# Towards Weyltronics: Realization of epitaxial NbP and TaP Weyl Semimetal thin films

*Amilcar Bedoya-Pinto\*, Avanindra Kumar Pandeya, Defa Liu, Hakan Deniz, Kai Chang, Hengxin Tan, Hyeon Han, Jagannath Jena, Ilya Kostanovskiy and Stuart Parkin\**

Max Planck-Institute of Microstructure Physics, Weinberg 2, 06120 Halle (Saale), Germany

*abedoyapinto@mpi-halle.mpg.de , *stuart.parkin@mpi-halle.mpg.de,

## ABSTRACT

Weyl Semimetals (WSMs), a recently discovered topological state of matter, exhibit an electronic structure governed by linear band dispersions and degeneracy (Weyl) points leading to rich physical phenomena, which are yet to be exploited in thin film devices. While WSMs were established in the monopnictide compound family several years ago, the growth of thin films has remained a challenge. Here, we report the growth of epitaxial thin films of NbP and TaP by means of molecular beam epitaxy. Single crystalline films are grown on MgO (001) substrates using thin Nb (Ta) buffer layers, and are found to be tensile strained (1%) and with slightly P-rich stoichiometry with respect to the bulk crystals. The resulting electronic structure exhibits topological surface states characteristic of a P-terminated surface, linear dispersion bands in agreement with the calculated band structure, and a Fermi-level shift of -0.2 eV with respect to the Weyl points. Consequently, the electronic transport is dominated by both holes and electrons with carrier mobilities close to $10^3$ cm$^2$/Vs at room-temperature. The growth of epitaxial thin films opens up the use of strain and controlled doping to access and tune the electronic structure of Weyl Semimetals on demand, paving the way for the rational design and fabrication of electronic devices ruled by topology.

**INTRODUCTION**

Weyl Semimetals, materials in which electrons obey the physics of Weyl fermions, are one of the most recent discoveries in topological matter [1,2]. The existence of Weyl Fermions was first predicted in materials breaking time-reversal symmetry such as pyrochlore iridates [3], but their experimental realization was ultimately achieved in compounds breaking inversion symmetry, the transition metal pnictides [4-12]. So far, these compounds were synthesized as bulk single crystals in the structural space group I4$_1$md and with composition MX (M=Nb,Ta, X=P,As). Their identification as Weyl Semimetals relied mostly on mapping the unique features in the electronic structure: a linear dispersion with band crossings (Weyl Points) and Fermi-arcs in the surface states have been conclusively measured by several groups using angle-resolved photoemission spectroscopy [6-11], and further confirmed by local spectroscopic techniques [13-16]. Moreover, as Weyl Fermions coupled to electromagnetic fields lead to non-conservation of chiral charges – the so called Adler-Jackiw anomaly [17] - its signatures were investigated in magnetotransport experiments, with some concern regarding the possible role of inhomogeneous current flow in bulk crystals [18-20]. Although major efforts have been put in the fabrication of WSMs micro-structures from bulk crystals by focused ion beam milling [21-24], thereby enabling advanced thermal transport experiments [21], fabrication-related property changes of the WSMs had to be taken into account. In this context, the realization of a clean bottom-up approach is an important step to unambiguously ascribe these anomalies to the intrinsic properties of the WSMs, making non-local transport experiments [25] a viable route. Furthermore, thin film growth would allow the precise control of the topological properties of WSMs via doping or strain, a yet unexplored terrain which can be barely accessed with the present bulk crystals. For instance, strain gradients in a WSM crystal are expected to induce electromagnetic gauge fields [26], leading to the observation of emergent phenomena such as landau levels and quantum oscillations in absence of a magnetic field [27], and a quantization of the circular photogalvanic effect [28], typically under strain conditions that can be well achieved by lattice-matched engineering during epitaxial growth. But the most important implication of WSM thin films is the possibility to fabricate atomically engineered heterostructures and functional interfaces. It is a unique opportunity to explore the interplay of WSMs with other materials, such as superconductors, (anti)ferromagnetic or ferroelectric materials. In this respect, there has been exciting predictions of interfacial effects that make WSMs appealing for spintronic and superconducting devices, such as a large topology-driven spin-Hall [29] and Edelstein effect [30], as well as a chirality-dependent Josephson current [31]. Moreover, there have been a number of fundamental phenomena in WSMs thin films addressed by theory, such as an unusual twisting of the Fermi surface [32], the emergence of Floquet topological insulator phases [33], a metal-insulator transition upon thickness confinement [34] or even a special interplay of long and short-range surface plasmon-polariton modes [35]. On the other hand, promising application areas of WSMs have been addressed, proving them as efficient hydrogen catalysts [36], colossal photovoltaic materials [37], mid-infrared detectors [38], and most recently, topological magnets [39-41]. It is thus clear that the fabrication of thin films will boost the impact of WSMs both in fundamental and applied research perspective.

In this article, we report the growth of high-quality NbP and TaP thin films on insulating MgO (100) substrates by molecular-beam epitaxy. Both phosphide compounds have been conclusively shown to be type-I Weyl Semimetals in the bulk crystal form by photoemission experiments [8-11]. Our epitaxial layers present clear differences with respect to the bulk crystals: (i) both in-plane and out-of plane lattice parameters are larger by more than 1%, (ii) the composition of the films is slightly in P-rich, (iii) the film surface exhibits sub unit cell steps (2.8 Å) that point to a single surface termination. The resulting electronic properties are marked by a downward shift of the Fermi-energy (0.2 eV) below the Weyl-points, which effectively increases the carrier density due to multiple band crossings. This effective hole-doping is likely to arise from the formation of Nb-vacancy acceptors during the P-rich growth process. A phosphorous-terminated surface is naturally achieved by the growth method and is confirmed by the shape of the electronic surface bands in both experiment and calculations, and allows for the observation of topological band dispersions. As for the electronic transport, although the Fermi-energy of the as-grown samples is still far from the Weyl points, electron mobilities close to $10^3$ cm$^2$/Vs have been measured at room temperature in patterned Hall-bar devices. The ability to grow thin films of Weyl semimetals that can be tailored by doping or strain, opens up new opportunities to use topology for electronic devices, thus setting the first milestone for the nascent field of Weyltronics.

**RESULTS AND DISCUSSION**

The NbP (TaP) compounds crystallize in the I4$_1$md structural space group, which is a tetragonal lattice with parameters a=3.34 Å (3.36 Å) and c=11.37 Å (11.41 Å). The method of choice for epitaxial growth is to use the basal (a-b) plane of the pnictide structure to match with substrates with a cubic structure, i.e. to grow the films along the (001) direction. However, for a reasonable lattice matching on typical cubic oxide insulators, the NbP (TaP) growth has to be stabilized with an in-plane rotation of 45° in the basal plane with respect to the substrate. In this regard, MgO (100) substrates were selected and prepared ex-situ to achieve an atomically flat surface prior to growth (more details about the substrate choice can be found in Supplementary Note 1). Besides the lattice matching issue, another challenge for the ultra-high-vacuum growth of phosphides containing refractory metals such as Ta or Nb, is to deal with the radically different vapor pressures of the compound elements, and in particular, with the suppression of undesired pyrolytic phosphorous species such as $P_4$ (see Methods section for a description of the optimized process).

Figure 1a summarizes the growth strategy with the help of in-situ reflection high energy electron diffraction (RHEED) patterns: (i) we start from an atomically flat MgO (100) surface, (ii) a subsequent growth of a thin Nb (Ta) (001) buffer layer, (iii) the phosphorization of the metal buffer layer and finally (iv) the overgrowth of the phosphide layer. This procedure results in a very streaky NbP (TaP) RHEED pattern indicative of a quasi layer-by-layer growth. From the analysis of the diffraction patterns, it is clear that the Nb (Ta) buffer layer grows 45° rotated with respect to the MgO substrate, following the epitaxial relationship MgO [100]// Nb [110] and MgO[110]//Nb[100] (see structural models in Fig. 1b for clarity). After completion of the buffer layer, the Nb surface is exposed to a controlled phosphorous atmosphere, until high-order streaks (twice the distance of the main streaks in reciprocal space) appear. This observation is consistent with the nucleation of a NbP monolayer that is shifted by a/2 with respect to the buffer layer, which is inherent to the structure of NbP along the growth direction (c-axis, see cross-section in Figure 1b). The surface phosphorization is another crucial step to guarantee a smooth NbP overgrowth: streaky RHEED patterns indicative of flat epitaxial NbP layers are observed reproducibly with the adopted growth strategy. It should be noted that periodic RHEED reflections are only visible in the high symmetry directions (45° periodicity), while there are no coherent patterns at intermediate angles, indicating that the NbP (TaP) films grow with a single-crystalline orientation without twinning/twisting of in-plane crystalline domains. *Ex-situ* X-ray diffraction measurements (Fig. 1c) corroborate the single-crystalline order: a sharp 90° periodicity in the film peaks is observed upon azimuthal rotation (Phi-scan). Furthermore, by plotting the Phi-scans of film and substrate together, it is shown that the epitaxial relationship (45 deg rotation in-plane) does not change throughout the whole film thickness.

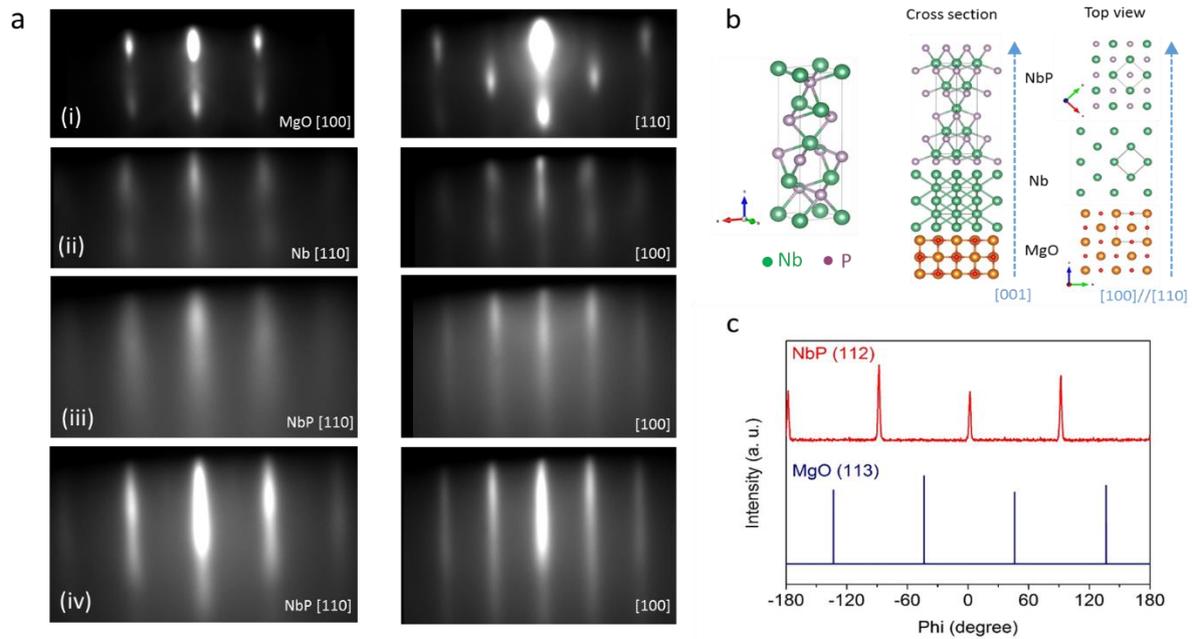

*Fig. 1: **Growth strategy and epitaxial relationships**. (a) Evolution of RHEED pattern during NbP epitaxial growth: (i) MgO substrate, (ii) Nb- buffer layer, (iii) Phosphorization of Nb surface and formation of first NbP monolayer (iv) NbP growth. (b) Sketch of NbP/TaP structure and epitaxial relationship between substrate, buffer layer and film. (c) X-ray diffraction Phi-scan of NbP (112) and MgO (113) reflection, indicating a four-fold symmetric, single-crystalline oriented growth rotated by 45° with respect to the substrate.*

Figure 2 shows a detailed investigation of the structural properties, comprising X-ray diffraction (XRD), high-resolution scanning transmission electron microscopy (S-TEM) and Raman-spectroscopy. A standard $\theta$-$2\theta$ scan taken on NbP films of various thicknesses (9-70nm) shows only (004) and (008) NbP reflections (Fig. 2a), confirming an epitaxial, single crystalline oriented growth without secondary phases. While the X-ray reflection pattern from the Nb buffer layer cannot be resolved due to the reduced thickness (2-5 nm), it can be well visualized in the overview cross-sectional TEM image in Figure 2b. As anticipated in the RHEED patterns during growth, the Nb buffer layer has a certain degree of structural disorder, which is not present in the NbP layers grown on top. The fine details of the structure are shown in Figure 2c, where a high-resolution high-angle annular dark field (HAADF) STEM image of the NbP in the [110] direction displays highly-ordered in-and out-of-plane lattice planes, with an excellent matching to the NbP structural model. The lattice planes are particularly visible by the high atomic contrast of the Nb atoms, while a faint contrast corresponding to the light P atoms can be distinguished in the zoomed-in image (Figure 2c, right panel) at the expected atomic positions. Taking advantage of the excellent structural order, a line intensity profile of the atomic rows has been taken to calculate the average in-plane lattice parameter of the NbP film (Supp. Figure S1). For the [110] direction, a Nb-Nb atom distance of 2.40 Å has been inferred, which results in an in-plane lattice parameter a=3.394 Å. In order to compare the quantification using both local and global methods, a reciprocal space mapping (RSM) scan has been performed on the (1,1,10) reflection of NbP (Fig. 2d), yielding an in-plane lattice parameter of a=3.391 Å, an excellent agreement within the error of the TEM linescan averaging and the determination of the XRD intensity maxima. The in-plane lattice parameters do not vary from as the thickness is varied from 15nm to 70nm according to the RSM measurements (see Sup. Fig S2), which means that the films do not change the in-plane strain state in the studied thickness range. On the other hand, the out-of-plane lattice parameters extracted from the (004) peak positions (inset of Fig. 2a) decrease only very slightly (11.50 Å to 11.46 Å) with increasing thickness. The negligible thickness dependence of both in-plane and out of-plane parameters suggest that the films grow fully relaxed from the very early stage. Interestingly, the lattice parameters (a=3.39 Å, c=11.48 Å) sizably differ from the bulk values (a=3.34 Å, c=11.37 Å), which means that a bigger unit cell is stabilized during the layer-by-layer growth on the MgO(001)/Nb(001) surface. Figure 2e shows the thickness-dependent Raman spectra of NbP thin films. Besides the elastic peak, the characteristic vibrational modes ($E_1$, $B_1$, $E_2$, $A_1$) are observed, which correspond to the NbAs structure type (space group $Im4c$). The most prominent peak belongs to the $A_1$ mode (around 380 cm$^{-1}$) while the other modes present a much lower intensity, in agreement with the reported bulk spectra and theoretical calculations [42]. We observe a small redshift of the $A_1$ mode as the thickness of the films increases (inset of Fig. 2e), which arises due to the slightly smaller c-lattice parameter found on thicker films. While in Ref. [42] the redshift was measured by applying hydrostatic pressure, in our case it was achieved by the thickness-dependent epitaxy. Thus, the shift of the $A_1$ mode can be used as a quick tool to probe the strain in the films, an important parameter for tuning the electronic properties of WSMs. Besides, unlike X-ray diffraction, Raman scattering is able to map different fingerprints for the isostructural NbP and TaP compounds, as evidenced in the shift of the $B^1$ mode (Supp. Figure S3).

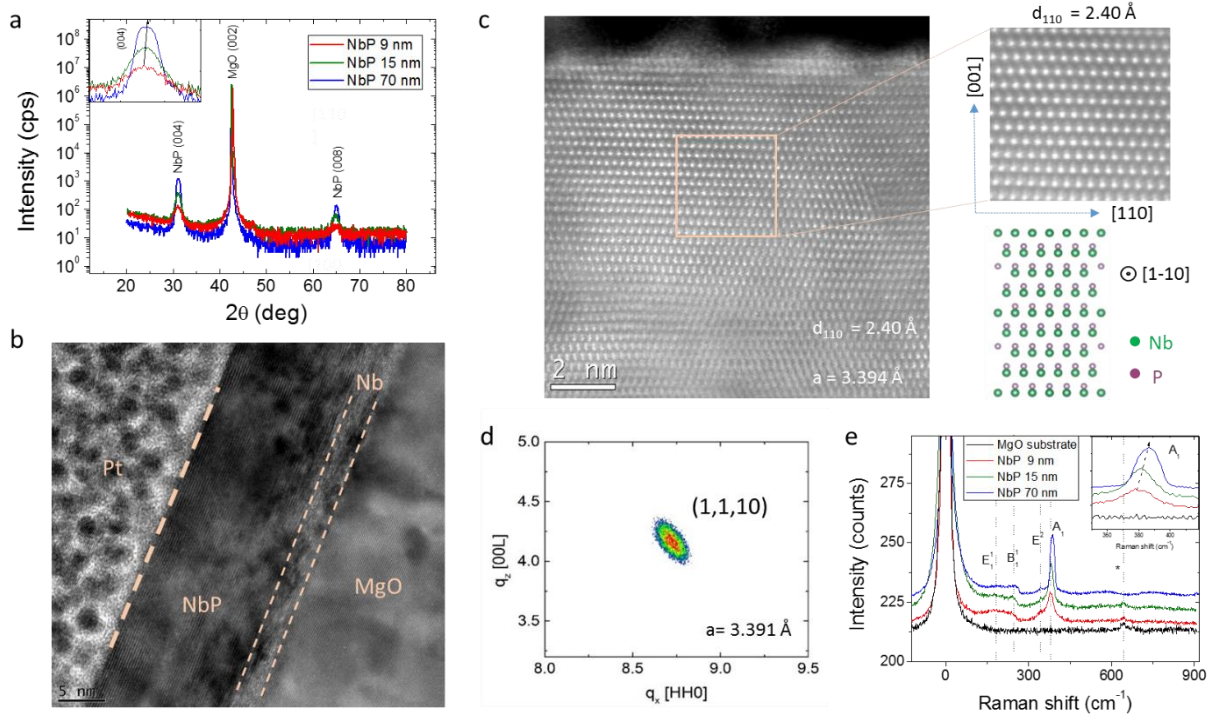

*Fig. 2:* ***Structural characterization by local and global probes***. *(a) X-ray diffraction (θ-2θ scan) of NbP thin films, indicating a (001) oriented film with the absence of secondary phases. (b) Transmission electron microscopy overview image of a MgO/ Nb(2nm) /NbP(15nm) cross-section. A Pt protection layer has been placed during TEM sample preparation. (c) High-resolution STEM images of the NbP film region in [1-10] direction, resolving the crystal structure and the location of the Nb and P atoms in the lattice. (d) Reciprocal space mapping of the (1,1,10) reflection of the NbP film studied by TEM. The extracted in-plane lattice*

*parameters match with the local TEM determination, confirming the large-scale homogeneity of the films. e) Thickness dependent Raman-spectra of NbP thin films. The shift of the $A_1$ mode towards higher frequencies is highlighted as inset.*

The chemical composition and valence states of the NbP (TaP) thin films have been studied by *in-situ* X-ray photoemission spectroscopy (XPS). The details of the Nb 3d 5/2 and P 2p core level spectra (Figure 3a-b) show that Nb and P shift in opposite directions compared to the neutral $Nb^0$ and $P^0$ valence state, consistent with the expected chemical shifts due to electron transfer in the NbP compound ($Nb^{III}$ and $P^V$ valence). The stoichiometry of the films has been determined using the area under the curve of the core levels with the respective sensitivity factors, yielding a slightly P-rich ($Nb_{0.49}P_{0.51}$) composition. Complementary to XPS, which is a surface sensitive technique, the in-depth composition of the films has been studied by Rutherford backscattering spectroscopy (RBS). The best fitting to the spectra (Figure 3c) yields a 47.6% to 52.4% (Nb:P) composition, indicating that the P-rich composition is distributed homogeneously across the full NbP layer. Similar results with regard to bonding-related core-level shifts and P-rich composition have been found in TaP layers (Supp. Figure S4). A real-space visualization of the layer homogeneity is further shown by energy-dispersive X-ray spectroscopy along a 15 nm NbP cross section (Figure 3d).

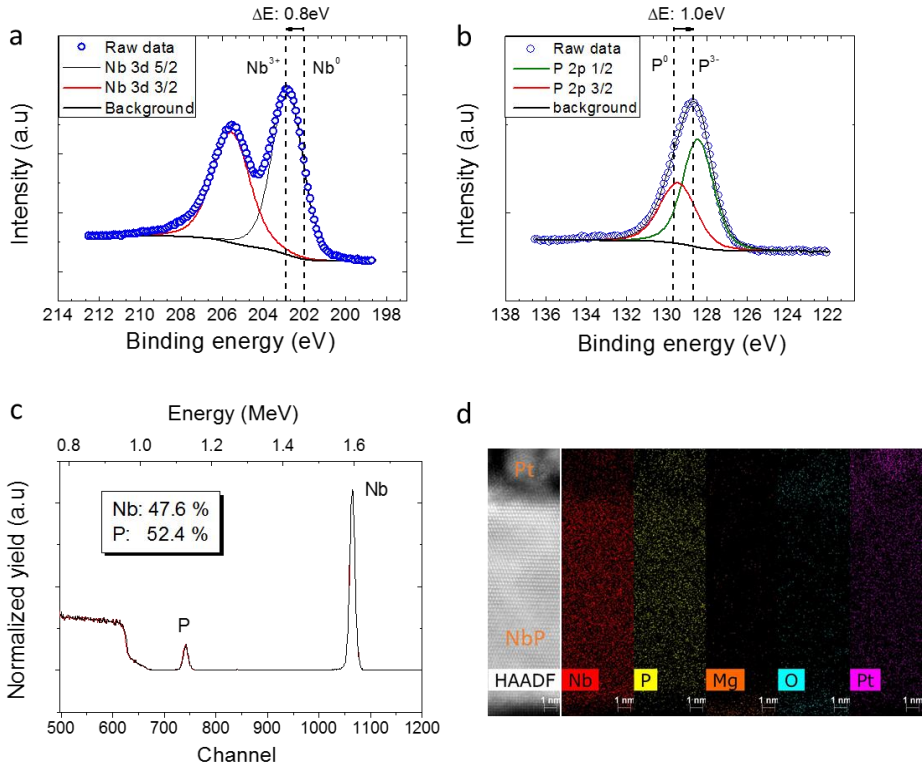

*Figure 3: Chemical and compositional analysis. (a,b) In-situ X-ray photoelectron spectroscopy of Nb 3d and P 2p core levels of NbP, highlighting the energy shifts in opposite directions due to chemical bonding. c) Rutherford backscattering spectra of a 20nm thick NbP sample. The red curve is the fit for composition determination. (d) Energy dispersive X-ray spectra on a TEM cross section, showing a homogeneous distribution of Nb and P species along the film thickness.*

The topography and surface structure of NbP and TaP films was investigated by scanning probe microscopy (Figure 4 and Supp. Fig S5). Large-scale atomic force microscopy images (Supp Fig. 5a) reveal a flat topography, yielding root mean square (RMS) roughness values of 0.43 nm. The grain size varies from 50 to 100 nm showing regions of grain coalescence, whereas the inter-grain steps correspond mostly to 1unit cell height. In order to investigate the topography inside and between the grains, *in-situ* scanning tunneling microscopy images have been acquired (Figure 4). Square-and rectangular shaped grains can be identified with two preferred orientations, along 45° and -45° on the image axis (i.e. along (100) direction), consistent with the four-fold in-plane crystal symmetry of NbP. A closer look to the grain topography reveals the presence of atomically flat terraces. The height of each step terrace amounts to 2.8 Å, corresponding to ¼ unit cell fractions (a single Nb-P monolayer), as depicted in Figure 2b (a structural model of the NbP unit cell along the growth direction (001) is drawn as guide to the eye). We rarely find slight deviations of exact unit cell fractions, which might arise if adjacent grains end in a different (Nb / P) atomic termination or due to the presence of inter-grain stacking faults. From the topography statistics, a predominantly single surface termination (either Nb or P) scenario is likely to happen throughout the film surface. Figure 4c and 4d show a zoomed-in topography image comparing NbP and TaP film surfaces, the latter having a smaller terrace width. Additional scanning tunneling spectra (STS) have been acquired on a flat region (marked with a square in Figure 4d) and reveal a parabolic behavior with a finite density of states at zero-bias, characteristic of metallic systems. An enhanced conductance (shoulder) is observed around + 0.2 V bias voltage, which points to a larger electronic density of states above the Fermi-energy (unoccupied states). The overall bias asymmetry of the dI/dV spectra is typically found for electronic systems lacking particle-hole symmetry and is not necessarily a feature arising from the density of states.

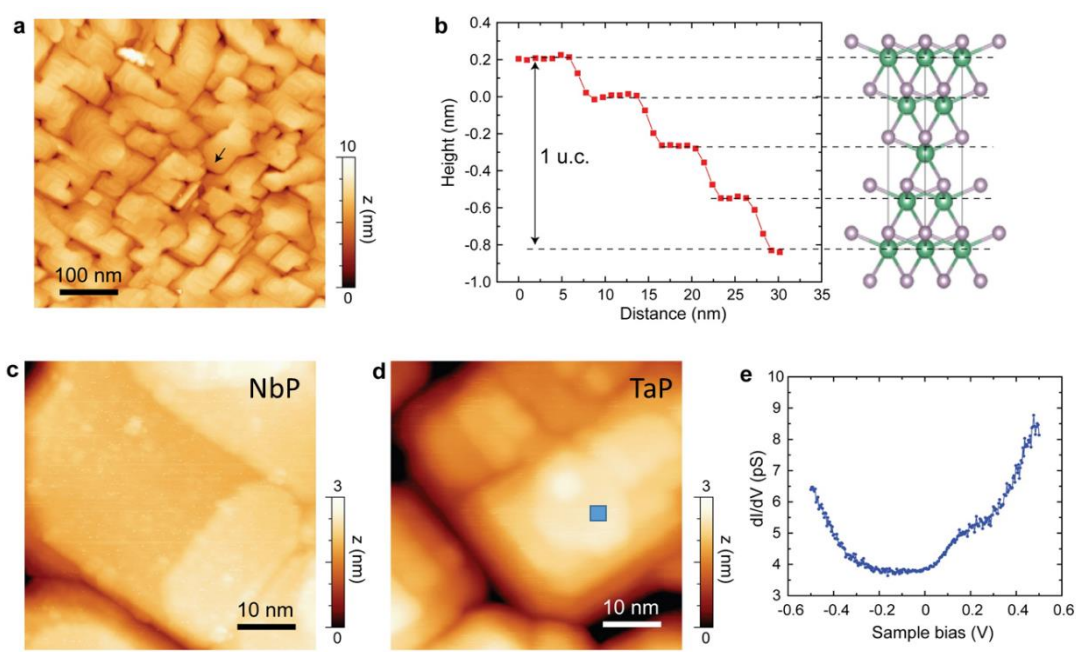

*Figure 4: **Surface properties by scanning tunneling microscopy and spectroscopy.** (a) Overview scanning tunneling microscopy image (500nm x 500nm, V = 1 V, I = 2 pA) of a 20nm-thick NbP film, revealing rectangular grains oriented along the (100) direction (45° in the image) with atomically flat terraces. (b) Height profile across a NbP grain (marked with an arrow in panel a), featuring terrace steps of 2.8 Å, which correspond to a NbP monolayer (1/4 u.c.). The NbP unit cell in (001) direction is drawn next to it as guide for the eye. (c,d) Zoomed-in topography images of NbP (V = 1 V, I = 2 pA) and TaP (V = 0.3 V, I = 10 pA), respectively. The terrace size in the TaP films appears to be smaller. (e) Scanning tunneling spectrum (STS) (V = 0.5 V, I = 10 pA) on a TaP epitaxial grain (marked as a square in panel d) reveals a finite density of states at zero-bias and a parabolic behavior with an enhanced conductance around 0.2V.*

Having assessed the properties of the film surface, momentum-resolved photoemission spectra have been taken using an in-house designed and built momentum microscope [43] with a He-I light source (see Methods for details). Figure 5 summarizes the overall electronic structure of a 15 nm-thick NbP thin film measured at 100K, including constant energy contours at different binding energies, band dispersion along relevant (Weyl point) cuts, together with ab-initio calculations. At the Fermi-energy, four electronic pockets with an elliptic shape directed towards the X and Y symmetry points can be identified. With increasing binding energy, these elliptical pockets evolve to a cross shape (200 meV) and finally the cross opens into an arc-like shape (400 meV), as shown in Figure 5a. Detailed analysis of this evolution and their comparison with the calculation of the termination-dependent surface states (Figure 5b and S6) reveal that the elliptic and cross-like band features along Γ-X and Γ-Y can be attributed to the P-terminated NbP surface states (the features for a Nb-termination are radically different, as shown in Supp. Figure S6). Interestingly, the size and shape of the measured elliptical (also called spoon-like) features match well with the calculations and the bulk crystal data in Refs.

[10,11,44] when an energy shift of $\Delta E = -0.2$ eV is considered (highlighted in a green box, Figure 5b), resulting in an effective hole doping in the as-grown thin films. Figure 5c shows the energy dispersion cut along A→A', which is expected to cross the location of one pair of Weyl-points in NbP ($k_x=0.54$ Å$^{-1}$, $E_b = -0.026$ eV) and thus used to visualize the surface Fermi arcs [10-11]. A clear linear band dispersion is observed, in agreement with previous photoemission results of cleaved monopnictide bulk crystals [6-11]. This dispersion originates from the Fermi-arcs, but their $k_x$-$k_y$ contour at the Weyl points cannot be mapped in our NbP films due to the $E_F$ shift (-0.2eV), and secondly, due the intrinsically short separation of the Weyl points in momentum space ($\Delta k < 0.05$ Å$^{-1}$), a challenge even for high-resolution synchrotron ARPES [10,11,44]. However, at deeper binding energies ($E_b = 400$meV, Fig. 5a) the observed open arc-like contours evolve from the Fermi-arcs and have thus topological character. The fact that only these topological surface states are observed in our films merits further investigation and will be discussed elsewhere [45]. The Fermi-level shift ($\Delta E$) of the epitaxial films is further visualized in the second derivative plot of the A→A' (panel (ii) in Figure 5c), considering the extrapolation to the band crossing point and the comparison to the shape of the bulk crystal dispersion data (inset of Figure 5c (ii)), taken from Ref. [11]. The origin of the effective hole-doping in the as-grown thin films can be mainly attributed to residual acceptors (Nb-vacancies) arising from the MBE growth process, in agreement with the slightly P-rich composition of the films inferred by XPS and RBS. However, the strained lattice parameters of the phosphide thin films will also have an effect on the Fermi-level determination. Although we estimate that the effect of the lattice parameters (~1% tensile strain) is rather small, further studies are needed to disentangle the contribution from strain and vacancy acceptors on $E_F$, in order to get a full understanding of Fermi-level engineering of Weyl semimetals achieved by epitaxial design.

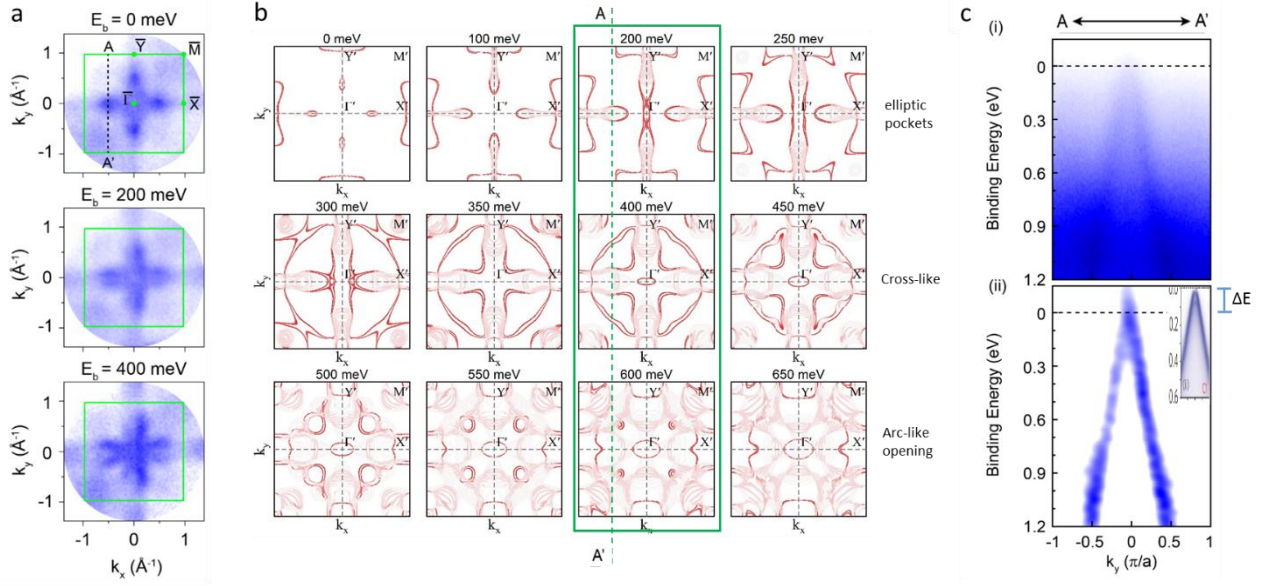

*Figure 5:* **Electronic structure of NbP thin films (t= 15nm) revealed by momentum-resolved photoemission spectroscopy.** *(a) Constant energy contours at different binding energies (0 - 400 meV). The green square represents the Brillouin zone (BZ). (b) Ab-initio calculation of constant-energy contours (surface projection) at different binding energies for a P-terminated surface. The calculated energy contours with an energy shift of ΔE = -0.2 eV (highlighted in green) match well with the experimental results in (a). (c) Energy dispersion cuts along the A-A' direction (defined in panel a), which corresponds to the cut across the Weyl point projection ($k_x=0.54$ Å$^{-1}$) [11,12], and its second derivative plot, respectively. A linear dispersion is clearly observed. Same cut taken from Ref. [11] is plotted and scaled as an inset for comparison.*

Electrical transport measurements have been carried out only for thick films (t = 70nm), so that shunting effects of the thin buffer layer can be discarded. The temperature dependence of the resistivity (Figure 6a) shows the expected metallic behavior for NbP. We do not observe superconductivity down to 2K, in contrast to recent reports on platelets formed from bulk crystals using focused ion beam milling [24]. The room-temperature resistivity (78 μΩcm$^{-1}$) is similar to values reported in bulk crystals (73 μΩcm$^{-1}$) [46] but much lower than values of FIB-fabricated microstructures (800 μΩcm$^{-1}$ to 4000 μΩcm$^{-1}$) [22,24], which means that there is no sample degradation upon micro-structuring in our epitaxial films. The carrier density in the as-grown films has been extracted by standard Hall-Effect measurements, and as already anticipated from the Fermi-level shift determined by photoemission ($\Delta E$= -0.2eV), both electron and hole bands will contribute to electronic transport. Thus, slightly non-linear transverse resistances ($R_{xy}$) as a function of magnetic field have been fitted by the two-carrier model to extract the densities of electrons ($n_e$) and holes ($n_p$), as well as the electron ($\mu_e$) and hole ($\mu_p$) mobility as a function of temperature, summarized in Figures 6b and 6c. Being energetically 0.2eV below the Weyl points, the density of states at $E_F$ crosses mainly hole-like bands, as shown in the bulk band structure calculations (Figure 6e), and thus the majority carriers are holes with a higher carrier density ($10^{21}$ – $10^{22}$ cm$^{-3}$) than in the bulk crystals ($10^{20}$ cm$^{-3}$) [46]. The resulting electron (hole) mobilities are close to 900 cm$^2$/Vs (300 cm$^2$/Vs) and feature a weak temperature dependence similar to heavily-doped (degenerate) semiconductors. Moreover, a positive magnetoresistance is observed both for in- and out-of plane magnetic fields (MR: 38% at 9T and 2K), evidencing the absence of chiral anomaly in the as-grown films. This result corroborates that the chiral anomaly is highly sensitive to the location of the Fermi-energy ($E_F$ = -0.2eV), since the effect of chiral charge pumping is strongly diminished away from the Weyl points, triggered by the contribution of non-topological bands at other locations in momentum space (see Figure 6e). At high magnetic fields, a linear, non-saturating behavior of the resistivity sets in (Supp Fig. S7), typically observed in high-mobility compensated semimetals.

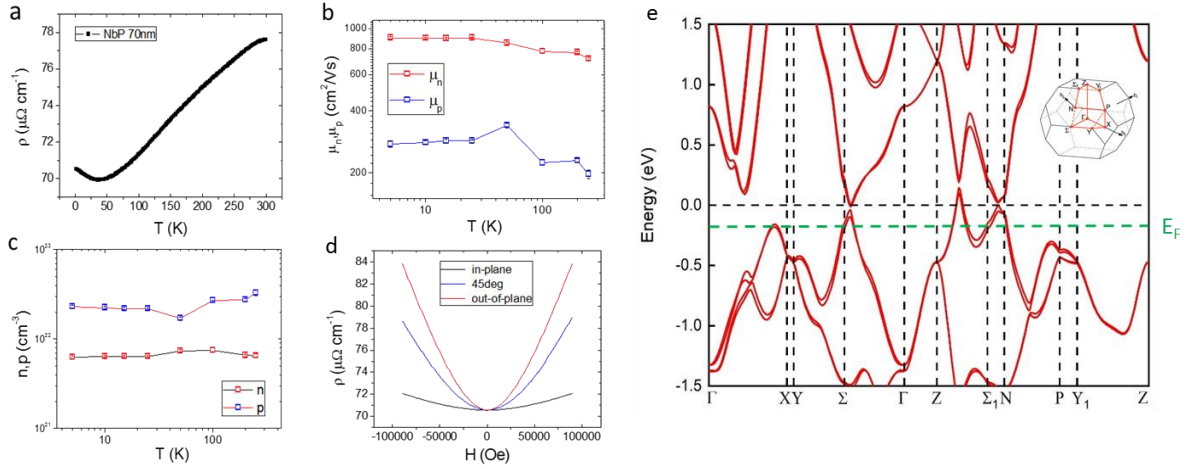

*Figure 6: Electrical and magnetotransport properties and its relation to the bulk bandstructure.* (a) Temperature dependence of the resistivity of a 70nm-thick NbP film, thickness at which the parallel buffer layer conduction (t=2nm) is negligible. (b) Electron and hole mobility calculated using the two-carrier model, exhibiting a very shallow temperature dependence. (c) Carrier density dependence reveals hole carriers as majority carriers throughout the measured temperature range. (d) Angular dependent magnetoresistance, exhibiting positive MR dependence both under in and out-of-plane fields (absence of chiral anomaly). (e) Bulk bandstructure calculations of NbP showing all the relevant bands inside the Brillouin zone, and highlighting the presence of multiple hole pockets at the experimentally determined $E_F$ (green line), in agreement with the large hole concentration ($>10^{22}$ $cm^{-3}$) obtained by Hall-Effect.

In summary, epitaxial thin films of type-I Weyl Semimetals NbP and TaP have been synthesized via molecular beam-epitaxy. A rigorous structural, morphological and chemical characterization was carried out to assess the quality of the films. Due to the epitaxial growth, the layers are stabilized with a larger lattice constant (1%) with respect to the bulk crystals and with a slight P-rich stoichiometry. The excellent surface quality, featuring a homogeneous P-termination, allows the visualization of linear dispersive topological bands. A Fermi-level shift of around -0.2eV with respect to the intrinsic $E_F$ is determined by comparing the photoemission spectra with ab-initio calculations and bulk crystal data, and arises mainly due to the formation of Nb-vacancy acceptors (P-rich conditions). The realization of high-quality WSM thin films and the intimate relation between strain, doping, band structure and electronic transport discussed here, opens up a route to access and control topological bands on demand, and paves the way towards the achievement of novel topological heterostructures and the first functional Weyltronic devices.

**METHODS**

**Substrate preparation and molecular-beam epitaxy growth.** MgO (100) substrates were soaked in methanol for 20 min and rinsed with water followed by annealing in $O_2$ atmosphere (1150°C) for 3.5 hours to achieve an atomically flat surface prior to film growth. The typical substrate dimensions are 5x10 mm. Nb (Ta) rods are evaporated via electron-beam heating and P species are thermally evaporated from a GaP compound effusion cell in a custom-made UHV chamber ($p_{base}$ = 1x10$^{-10}$ mbar), with a regeneration system for residual $P_2$ (red) and $P_4$ (white phosphorous). The GaP compound cell is employed to reduce the amount of white phosphorous upon evaporation ($P_2/P_4 \sim 100$). A cross-beam mass spectrometer (XBS Hiden) is used to calibrate the atomic fluxes and monitor the amount of $P_2/P_4$ species. No Ga-species have been detected under evaporation conditions ($T_{cell}$ = 850°C). The substrate temperature is controlled by radiation heating. The Nb (Ta) buffer layer is grown at 300°C, at a rate of 3-5nm/h and a pressure of p= 4 x10$^{-10}$ mbar. The surface of the buffer layer is then exposed to a $P_2$ flux (BEP: 1 x10$^{-8}$ mbar) to achieve phosphorization, whereas the subsequent NbP (TaP) layer is grown under P-rich conditions (Nb:P flux 1:20), a moderate substrate temperature (300-400°C) and a slow rate (< 4 nm/h), controlled by the Nb(Ta)-flux. The thickness of the films presented in this work ranges between 9 and 70nm. After concluding the growth process, the sample is cooled down very slowly (10°C/min) to room-temperature under P-atmosphere (p= 1 x$10^{-8}$ mbar), to ensure a homogeneous P-termination at the surface.

**In-situ characterization.** The crystallinity of the films is monitored in-situ by Reflection High Energy Electron Diffraction (RHEED), using a 15 kV electron beam. The layers are further characterized by in-situ tools (XPS, STM and momentum microscope) with the use of a vacuum suitcase transfer system (Ferrovac, $p_{base}$ < 1x$10^{-10}$ mbar) to move the as-deposited film under UHV conditions from the growth to the analysis chambers. Core-level X-ray photoelectron spectra (XPS) were taken at room-temperature using Al Kα radiation and a hemispherical analyzer. The STM experiments were performed on an Omicron VT-STM-XT system operated at room temperature with a base pressure of 2 x $10^{-11}$ mbar. The mechanically sharpened Pt/Ir tips were treated and checked on Au(111) surface before measurements, and the topography images were acquired at room-temperature. A sinusoidal modulation of 30 mV, 713 Hz was added to the bias voltage for the measurements of dI/dV spectra. Momentum-resolved photoemission spectra was acquired using an in-house designed momentum microscope [1], using a Helium lamp (hv=21.2 eV), at 100K and with an energy step of 10 meV.

**Ex-situ characterization.** X-ray diffraction patterns, including reciprocal space maps, were taken in a commercial 4-circle diffractometer (Bruker) using Cu Kα radiation. Transmission Electron Microscopy (TEM) was carried out in an aberration corrected high-resolution microscope (FEI, Titan). Cross-sectional lamellas of MgO/Nb/NbP stack (t = 100 nm) were prepared by Focused Ion Beam (FIB) with Ga-ion etching. A Pt protection layer is deposited on the sample before lifting the lamella to the TEM grid. Raman spectra was acquired using 532 nm excitation wavelength at room temperature. Rutherford Backscattering Spectroscopy using Particle Induced X-Ray Emission was measured using a 1.9 MV alpha+ beam at a 169° scattering angle in an NEC Pelletron accelerator.

**Device fabrication and electrical transport.** Hall-bar devices were patterned using laser-assisted optical lithography (Heidelberg Instruments) and etched via an Ar-discharge plasma (500 V DC) under high-vacuum conditions. Cr/Au (4 nm / 80 nm) contacts were deposited by DC magnetron sputtering. The devices size ranged from (5 x 15) µm$^2$ (width x length) to (100 x 300) µm$^2$ (aspect ratio 1:3).

**Ab-initio calculations.** Eelectronic structure calculations are performed within the density functional theory as implemented in the Vienna ab initio simulation package (VASP) [2]. The exchange-correlation energy is treated within the Perdew-Burke-Ernzerhof (PBE) parametrization [3] of the generalized gradient approximation. The kinetic energy cutoff for the plane wave basis is 300 eV and a k-mesh of 500×500×1 is adopted for the Brillouin zone integration. A six-unit-cell thick slab model is constructed by cutting NbP along the (001) plane of the conventional cell and the in-plane cell size is 1×1 conventional cell. The internal coordinates of the atoms in the surface unit cell are fully relaxed. Spin-orbit coupling is considered in the electronic structure calculations. All the surface band structures and Fermi surfaces are projected onto the atoms in the surface unit cell of the P- and Nb-termination, accordingly.

**Supporting Information**

Supplementary Information available

**Author Contributions**

A.B-P and S.S.P. conceived the project. A.B-P customized the MBE system, developed the film growth strategy, prepared the films and performed sample characterization and data analysis (RHEED, XRD, Raman, Electrical transport). A.K.P assisted with the film growth, carried out XPS measurements, device fabrication and data analysis (XPS, 2-carrier model). D.L. performed the momentum-resolved photoemission measurements and data analysis. H.T. calculated the bulk band structure and termination-dependent surface states. H.D. carried out the TEM measurements and J.J. prepared the cross-section lamella by FIB. K.C. carried out the STM/STS measurements and data analysis. H.H. performed the RSM measurements / Phi-scans and data analysis. I.K. carried out the RBS measurements and data analysis. A.B-P. wrote the manuscript with contributions of all of the authors. S.S.P. supervised the entire project.

**ACKNOWLEDGMENT**

We thank G. Woltersdorf and C. Koerner (MLU Halle) for their assistance regarding Raman measurements, as well as Y. Sun (MPI-CPfS) for theoretical input. D.L. thanks the Alexander von Humboldt Foundation for support.